\documentstyle[aps,prl,twocolumn]{revtex}

\begin{document}

\draft

\title{Dynamic Fano Resonance of Quasienergy Excitons in Superlattices}

\author{Ren-Bao Liu and Bang-fen Zhu}

\address{Center for Advanced Study, Tsinghua University, Beijing 100 084,
 People's Republic of China}

\maketitle

\abstract{ The dynamic Fano resonance (DFR) between discrete
quasienergy excitons and sidebands of their ionization continua
is predicted and investigated in dc-
and ac-driven semiconductor superlattices. This DFR, well
controlled by the ac field, delocalizes the excitons and opens
an intrinsic decay channel in nonlinear four-wave mixing signals.}


\normalsize

The Fano resonance (FR) results from quantum coupling
between a discrete state and a degenerate continuum of
states and manifests itself
in optical spectra as asymmetric lineshape\cite {Fano}.
FR was observed in a variety of atomic and molecular
systems. Recently, this phenomenon has been reported
in semiconductor quantum wells \cite{FRQW}, in biased
semiconductor superlattices (SSL's) \cite{FRSSL}, and
in bulk GaAs in the presence of magnetic field
\cite{FRMag,FRFWM} where discrete excitons couple
to continua of the lower transitions through Coulomb
interaction \cite{Review}.
In addition to the frequency-domain experiments, transient
four-wave mixing (FWM) in magneto-confined bulk
semiconductors\cite{FRFWM} and in SSL's\cite{FRSSL} has
demonstrated that the FR is a fundamental type of dephasing
mechanism in irreversible quantum dynamical processes.
 
An intense ac field, e.g. the free electron laser, can induce
the dynamical localization (DL) \cite{DL} of
electrons in SSL's. The DL is directly related to the collapse
of the quasienergy band of Floquet states
\cite{collapse1,collapse2,collapse3},
the temporal analogue to the Bloch states
in spatially periodic potential \cite{Floquet}.
Research taking into account Coulomb interaction has shown that the DL may
cause the dimension crossover of excitons \cite{XDL} and the enhancement
of exciton binding and opacity in linear \cite{XDL,XDL1} and nonlinear
spectra \cite{DDL}. Meanwhile the dynamical delocalization (DDL) effect
of an ac field is also investigated on localized Wannier-Stark (WS)
states \cite{WSL} in biased SSL's \cite{collapse2,collapse3,DDL},
which is hindered by Coulomb binding \cite{DDL}.

In this Letter, we predict the dynamic Fano resonance
(DFR) between quasienergy exciton states in biased SSL's
driven by an intense THz-field.
In particular, the novelty of this DFR consists in the
coupling between the discrete quasienergy exciton and
the {\em neighbor} sideband of its ionization continuum.
This DFR will also lead to the
dynamical ionization of bound excitons in SSL's, which itself
is a new effect of the ac field.

The one-dimensional tight-binding model \cite{DDL,FSBE} is adopted
in the present investigation. With excluding the realistic
three-dimensional excitonic motion, the results are hardly compared
 to experimental data quantitatively, but we believe this simple model
does present qualitatively correct results, in view of the fact that the
ground WS exciton-state usually possesses much larger oscillator
strength than its ionization continuum of in-plane motion, and the
coupling induced by the ac field between the in-plane motion associated
with different WS states is negligible because of their approximate
orthogonality. The Coulomb interaction is treated on the basis of
Hartree-Fock approximation. The neglected genuine many-body
correlation \cite{Axt} may lead to almost instantaneous decay in
time-integrated (TI) FWM signals, which is irrelevant to the intrinsic
dephasing \cite{FRFWM}.
This effect is crucial for the magneto-confined FR \cite{FRFWM} due to
the magnetic field-induced enhancement of many-body correlation \cite{MagQK},
but is less important for the decay of TI-FWM signals in biased SSL's, which
is, in essence, an intrinsic dephasing process \cite{FRSSL,Zener}.

The semiconductor Bloch equations in the presence of electric fields
\cite{FSBE} for the exciton amplitude $p_k\equiv\langle a_{{\rm h}-k}
a_{{\rm e}k}\rangle$ and electron-hole (e-h) pair density
$f_k\equiv\langle a^{\dag}_{{\rm e}k}a_{{\rm e}k}\rangle=
\langle a^{\dag}_{{\rm h}-k}a_{{\rm h}-k}\rangle$ read
\begin{eqnarray}
\partial_tp_k=&&-i\left[E_0+\varepsilon_k(t)\right]p_k+
\left[\omega_{\rm BO}+\omega_1\cos(\omega t)\right]
\partial_kp_k \nonumber \\ &&
+i\chi^R(t)(1-2f_k)-\gamma_2p_k, \label{exciton}\\
\partial_tf_k=&&\left[\omega_{\rm BO}+\omega_1\cos(\omega t)\right]
\partial_kf_k-2\Im\left\{\chi^R(t)p_k^*\right\}, \label{population}
\end{eqnarray}
where $a_{{\rm e}k}$ ($a_{{\rm h}k}$) is the electron (hole) annihilation
operator for the quasi-momentum state $|{\rm e(h)}, k\rangle$, $E_0$ is
the separation between centers of the conduction and the valence minibands,
$\varepsilon_k(t)=\varepsilon_k-2\sum_{k'}V_{k,k'}f_{k'}$ is
the renormalized combined e-h miniband dispersion
with $\varepsilon_k=-\frac{\Delta}{2}\cos k$,
$\omega_{\rm BO}$ is the Bloch oscillation frequency due to the dc-field
\cite{BO}, $\omega_1$ is the strength of the driving ac field with frequency
$\omega$, $\chi^R(t)=\chi(t)+\sum_{k'}V_{k,k'}p_{k'}$ is the renormalized
strength of a near infrared optical excitation (here the dipole element is
assumed $k$-independent), and $V_{k,k'}$ comes from the Coulomb potential.

The Schr{\"o}dinger equation for the exciton wave function is just the
linear part of Eq. (\protect\ref{exciton}) with the excitation and dephasing
terms removed. The quasienergy is obtained by numerically diagonalizing
the propagator
$$U(T+t,t)\equiv \hat{\rm T}\exp\left(-i\int^{T+t}_tH(t)dt\right)$$
[$H(t)$ is the Hamiltonian, $T\equiv 2\pi/\omega$], according to
the secular equation
$$U(T+t,t)u_q(k,t)=\exp\left(-i\varepsilon_qT\right)u_q(k,t),$$
where $u_q(k,t)$ is the
eigenstate with quasienergy $\varepsilon_q$. Obviously, for any
integer number $m$, $u_{q,m}\equiv u_q\exp(im\omega t)$ is also a
solution with quasienergy $\varepsilon_q+m\omega$, and, in analogy
to the conventional Brillouin zone in quasi-momentum space\cite{Floquet},
so is defined the Brillouin zone of
quasienergy $\left\{[m\omega,(m+1)\omega]\right\}$.

In the ac-driven system, the absorption spectrum $\alpha (\Omega)\propto
\Im\left[P^{(1)}(\Omega)/\chi(\Omega)\right]$,
where $P^{(1)}(\Omega)$ is the linear term of the Fourier
transformation of the optical response $P(t)\equiv \sum_k p_k$,
and $\chi(\Omega)$ is the excitation profile \cite{DDL1,Jauho1}.
In this work, only the continuous wave absorption is considered, whereby
the interband susceptibility and oscillator strength are of the same
form as in the static case, except that the transition matrix element
is replaced by its time-average and the transition energy by the
quasienergy \cite{Jauho1}. The FWM signals are calculated following
Ref. \protect\ref{liu}, in which two degenerate Gaussian-shaped
pulses [$\chi(t)=\chi\exp(-i\Omega_0 t-G^2t^2/2)$] are separated
by a delay time $\tau$.

For the sake of simplicity, let us focus on a specific case: The ac field is
resonant with the Bloch oscillation, and the Coulomb potential is of on-site
type, i.e. $V_{k,k'}=V_0/N$ ($N$ is the size of the basis set
$\{|k\rangle\}$). Parameters for our calculations are taken as follows.
$V_0=10$ meV, $\Delta=4V_0$, $\omega_{\rm BO}=\omega=1.5V_0$,
$\gamma_2=0.1V_0$, $\Omega_0=E_0-V_0$, and $G=\omega_{\rm BO}$.
$N$ is chosen to be 40 for calculating quasienergy, and 80 for
optical spectra. Our calculation shows that a larger basis
produces no significant difference.

As shown in Fig. \protect\ref{energy}, the quasienergy spectrum consists of
continuous minibands and a few well-separated discrete states below. As the
ac field strengthens, owing to the DDL, the miniband broadens and the
discrete excited states \cite{note1} merge into the continuum one by
one. Meanwhile, the discrete ground state is repelled by its continuum
towards lower energy, and eventually dips from above into the ionization
continuum of its neighbor sideband, causing a series of anticrossing in
the spectrum. Further enhancement of the ac field may suppress
the quasienergy miniband, then the discrete exciton states can
be released from the miniband one by one. At certain strength, namely
$\omega_1=\omega_{\rm DL1}$, the miniband collapses and DL takes
place. After that the evolution described above will repeat with
increasing the ac field.

For an intuitive understanding, let us briefly review the picture of
dressed WS ladders introduced in Ref. \protect\ref{liu}. In weak
ac field limit, only one-photon-assisted hopping between
neighbor WS states $|n\rangle_{\rm x}$ takes effect \cite{DDL,DDL1},
thus the time-periodic states $\{\exp(-in\omega t)|n\rangle_{\rm x}\}$
form the Wannier basis set in a tight-binding `lattice' with the
nearest-neighbor hopping coefficient of
$\omega_1\Delta/(8\omega_{\rm BO})$ \cite{DDL}.
Those Wannier states with small $n$ can be viewed as `impurities'
in a crystal since their `on-site energy' $E^{\rm x}_n-n\omega$
deviates from zero significantly due to the
Coulomb interaction \cite{XWSL}. The discrete states in the quasienergy
spectrum are concentrated on these `impurities'; on the other hand, the
remote WS states are almost equally spaced and the resonant photon-assisted
hopping results in the formation of the continuum as wide as without
the Coulomb interaction. As the DDL is enhanced by the ac field, the
discrete excited states may be ionized, and merge into the continuum.
Calculation with this picture agrees well with the exact results for
$\omega_1<\omega/2$ (not shown). However, when the ac field is strong
enough, the inter-sideband interaction with multi-photon
processes\cite{DDL} has to be invoked to account for
the mixing between the discrete ground state and the
continuum, the band suppression, and the DL.

The absorption spectrum in the absence of an ac field presents just
the excitonic WS ladder (Fig. \protect\ref{absorption}, $\omega_1/
\omega_{\rm DL1}=0$). When the ac field is turned on, the WS states are
coupled through photon-assisted hopping, so their sidebands start to gain
in oscillator strengths \cite{DDL,DDL1}, which looks as if the WS peaks were
split (Fig. \protect\ref{absorption}, $\omega_1/\omega_{\rm DL1}=1/16$).
As the excited discrete states merge into the continuum, the
{\em intra-sideband} DFR occurs, and the oscillator strength is
shared among the continuum states which otherwise have negligible
opacity\cite{DDL}. This manifests itself in the spectrum as broadening
of the original discrete line (Fig. \protect\ref{absorption},
$\omega_1/\omega_{\rm DL1}=1/16$ and $1/8$). When the ground exciton
state couples to energetically degenerate continuum associated with
its neighbor sideband, (for $\omega_1/\omega_{\rm DL1}=1/8$, 1/4, 1/2,
and 3/4 in Fig. \protect\ref{absorption}), the {\em inter-sideband} DFR
happens, as characterized in the absorption spectra by asymmetric
peaks with FWHM larger than $\gamma_2$ . As shown in the inset of
Fig. \protect\ref{absorption}, these peaks have perfect Fano
lineshape \cite{Fano}
\begin{equation}
\alpha(\Omega)=\alpha_0+\alpha_{\rm c}
\frac{(q\gamma+\Omega-E_{\rm x})^2}{\gamma^2+(\Omega-E_{\rm x})^2},
\label{Fano}
\end{equation}
where $\alpha_0$ is the background constant, $\alpha_{\rm c}$
represents the continuum absorption without the
inter-sideband coupling, $q$ is the lineshape parameter,
$\gamma$ is related to the resonance broadening,
and $E_{\rm x}$ denotes the position of the discrete
state. When further enhancement of the ac field reduces
the quasienergy-band width and the discrete states emerge
out of the continuum, the absorption spectrum evolves
from the DFR to Lorentzian-shaped discrete lines.
 
The lineshape of the absorption peaks vary drastically when
the ac field strength passes through the field at which the DFR occurs.
As shown in Fig. \protect\ref{fit}, when the exciton ground state
meets the continuous sideband, $1/q$ becomes nonzero, demonstrating the
asymmetrical non-Lorentzian lineshape, and the Fano coupling
parameter, $\gamma$, is larger than the static dephasing rate of excitons, 
$\gamma_2$, indicating that the peak is further broadened by the DFR.
According to Ref. \protect\ref{fano},
$$\gamma=\pi|V_E|^2$$
(in the present calculation, $\gamma$ also includes the contribution
from $\gamma_2$), and
$${1}/{q^2}=\left|{\pi V_E}\left({\mu_E}/{\mu_X}\right)\right|^2,$$
where $V_E$ is the coupling matrix element between the discrete and
the continuum states, and $\mu_E$ and $\mu_X$ are the optical transition
matrix element of the continuum and the discrete states, respectively.
$V_E$ is determined mainly by the overlap integral of the bound and
extended excitons, which can be modulated by the ac field through
changing the localization property of the excitons. Besides, the
ac field redistributes the oscillator strength among the quasienergy
states and their sidebands, changing the relative opacity
of the continuum. Thus both $1/q$ and $\gamma$, depending
on the sideband index, are readily tuned by the ac field [see Fig.
\protect\ref{fit} (a) and (b)].

The TI-FWM traces in Fig. \protect\ref{TIFWM} display the exponential
decay superimposed with quantum beats, which result from the
interference between the quasienergy excitons with different transition
energy \cite{QB}. The effective dephasing rate $\gamma^{\rm eff}_{2}$,
extracted from Fig. \protect\ref{TIFWM} and plotted
in Fig. \protect\ref{fit} (c), follows approximately the Fano coupling
parameter $\gamma$  [Fig. \protect\ref{fit} (b)], indicating that the DFR plays
a role in intrinsic phase-breaking process as the usual FR does in biased
SSL's\cite{FRSSL,Zener}.

To study the localization property of excitons in the dc- and ac-driven
SSL's, by integrating the Schr\"{o}dinger equation in the Wannier
representation, we have calculated the mean-square root of the exciton radius
$W\equiv \langle \hat{r}^2\rangle^{1/2}$ and the probability $R_0$ of
finding the electron and hole at the same site, at various delay time after
a weak $\delta$-pulse excitation [see Fig. \protect\ref{fit} (d) and (e)].
A $\delta$-pulse excites the e-h pair at the same site, so at $t=0^+$, $W=0$
and $R_0=1$. $W$ increases unboundedly with time as a result
of the resonant photon-assisted hoping between the evenly spaced
remote WS states, whenever the quasienergy miniband has finite width.
This DDL effect is handicapped somewhat by the Coulomb binding: The $R_0$
may saturate with time whenever there remains a discrete
state outside the continua. When the inter-sideband DFR occurs,
however, the excitons are dynamically ionized with no such saturation
behavior for $R_0$.

In summary, the dynamic Fano resonance between a discrete exciton
and the sideband of ionization continuum is predicted in dc- and
ac-driven SSL's, which, well controlled by the ac field strength,
manifests itself as broadened asymmetric lineshape in absorption
spectra and intrinsic decay in FWM signals, and leads to the dynamic
ionization for bound excitons. All these effects stem essentially
from the unequal spacing of the excitonic WS ladder together with
the DDL effect of the ac field. This DFR effect was absent in
the absorption spectra calculated in Ref. \protect\ref{crossover},
because the sidebands of quasienergy miniband are not effectively
overlapped for the chosen parameters . It should be pointed out that
the DFR is quite different in nature from the FR between WS states
and the degenerate in-plane continua associated with low-lying WS
states in dc-biased SSL's \cite{FRSSL}. The DFR is expected to be
experimentally distinguishable from the in-plane FR in its ac field
dependence. Moreover, it's possible to adjust the dc-field strength
to minimize the in-plane FR effect \cite{FRSSL} so as to emphasize
the dynamic interference.

This work was supported by the National Science Foundation of China.

\begin{figure}
\caption{Quasienergy sideband structure of excitons as
 functions of the ac field strength.}
\label{energy}
\end{figure}

\begin{figure}
\caption{ Absorption spectra (solid lines) in the dc- and ac-driven
 superlattice at various ac field strength as indicated by
 $\omega_1/\omega_{\rm DL1}$. The integer numbers denote the WS state index.
 Inset is an enlarged example of the lineshape due to the DFR, where the
 squares are fitted with Eq. (\protect\ref{Fano}).
 The dotted line represents the excitation spectrum for
 calculating the FWM signals.}
\label{absorption}
\end{figure}

\begin{figure}
\caption{Dependence on the ac field strength of (a) the inverse
 lineshape parameter $1/q$ and (b) the Fano coupling parameter $\gamma$
 obtained by fitting with Eq. (\protect\ref{Fano}) the absorption peaks
 associated with the exciton ground state in Fig. \protect\ref{absorption},
 (c) the effective dephasing time $\gamma_2^{\rm eff}$ extracted from
 the TI-FWM signals, and (d) the mean-square root of the exciton radius $W$
 and (e) e-h overlap probability $R_0$ at $t=0^+$, $10T$, $20T$, $30T$,
 and $40T$ after a $\delta$-pulse excitation. In (a) and (b), the index
 `I$m$' denotes the exciton ground state in the quasienergy Brillouin zone
 $[(m-1)\omega,m\omega]$.}
\label{fit}
\end{figure}

\begin{figure}
\caption{TI-FWM signals as functions of delay time in the
 dc- and ac-driven superlattice for various ac field strength
 indicated by $\omega_1/\omega_{\rm DL1}$.}
\label{TIFWM}
\end{figure}

\begin{references}
\bibitem{Fano} U.Fano, Phys. Rev. {\bf 124}, 1866 (1961).\label{fano}
\bibitem{FRQW} D. Y. Oberli, G. B\"{o}hm, G. Weimann, and J. A. Brum,
 Phys. Rev. B {\bf 49}, 5757 (1994).
\bibitem{FRSSL} C. P. Holfeld {\it et al.}, Phys. Rev. Lett. {\bf 81},
 874 (1998).
\bibitem{FRMag} S. Glutsch, U. Siegner, M.-A. Mycek, and D. S. Chemla,
 Phys. Rev. B {\bf 50}, 17 009 (1994).
\bibitem{FRFWM} U. Siegner, M.-A. Mycek, S. Glutsch, and D. S. Chemla,
 Phys. Rev. Lett. {\bf 74}, 470 (1995).
\bibitem{Review} For a review of FR in semiconductors, see S. Glutsch,
 {\it Festk\"{o}rperprobleme/ Advances in Solid State Physics}, edited
 by R. Helbig (Vieweg, Braunschweig/ Wiesbaden, 1998), Vol. 37, p. 151.
\bibitem{DL} A. A. Ignatov and Y. A. Romanov, Phys. Status Solidi (B)
 {\bf 73}, 327 (1976); D. H. Dunlap and V. M. Kenkre, Phys. Rev. B
 {\bf 34}, 3625 (1986).
\bibitem{collapse1} M. Holthaus, Phys. Rev. Lett. {\bf 69}, 351 (1992).
\bibitem{collapse2} J. Zak, Phys. Rev. Lett. {\bf 71}, 2623 (1993).
\bibitem{collapse3} X.-G. Zhao, J. Phys. Condens. Matter {\bf 6}, 2751
 (1994).
\bibitem{Floquet} For general Floquet theory on time-periodic systems,
 see, e.g., J. H. Shirley, Phys. Rev. {\bf 138}, B979 (1965).
\bibitem{XDL} T. Meier, F. Rossi, P. Thomas, and S.W. Koch,
 Phys. Rev. Lett. {\bf 75}, 2558 (1995).\label{crossover}
\bibitem{XDL1} P. Ray and P. K. Basu, Phys. Rev. B {\bf 50}, 14 595
 (1994).
\bibitem{DDL} R.-B. Liu and B.-F. Zhu, Phys. Rev. B {\bf 59}, 5759
 (1999).\label{liu}
\bibitem{WSL} G. H. Wannier, Phys. Rev. {\bf 117}, 432 (1960).
\bibitem{FSBE}T. Meier, G. von Plessen, P. Thomas, and S. W. Koch,
 Phys. Rev. Lett. {\bf 73}, 902 (1994).
\bibitem{Axt} V. M. Axt and S. Mukamel, Rev. Mod. Phys. {\bf 70}, 145
 (1998), and references therein.
\bibitem{MagQK} P. Kner {\it et al.}, Phys. Rev. Lett. {\bf 78}, 1319 (1997).
\bibitem{Zener} B. Rosam {\it et al.}, Physica B {\bf 272}, 180 (1999).
\bibitem{BO} C. Zener, Proc. R. Soc. London Ser. A {\bf 145}, 523 (1934).
\bibitem{DDL1} M. M. Dignam, Phys. Rev. B {\bf 59}, 5770 (1999).
\bibitem{Jauho1} K. Johnsen and A.-P. Jauho, Phys. Rev. Lett. {\bf 83},
 1207 (1999).\label{Jauho1}
\bibitem{note1} Here the ground and excited states are designated according
 to their positions in a quasienergy Brillouin zone.
\bibitem{XWSL} M. M. Dignam and J. E. Sipe, Phys. Rev. Lett.
 {\bf 64}, 1797 (1990).
\bibitem{QB} J. Shah, {\it Ultrafast Spectroscopy of Semiconductors
 and Semiconductor Nanostructures}, (Springer, 1996), pp. 63-78, and
 references therein.
\end{references}
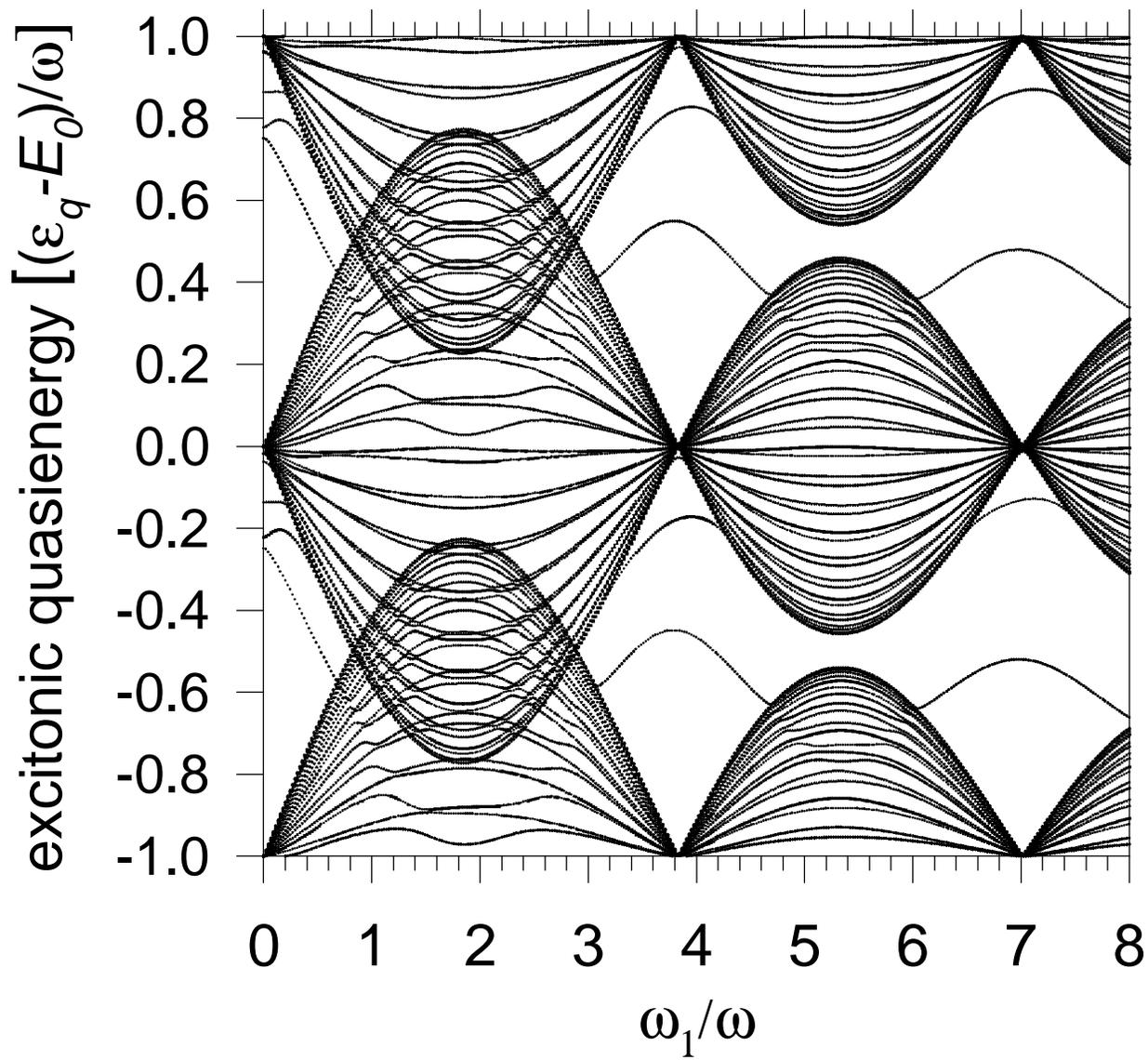
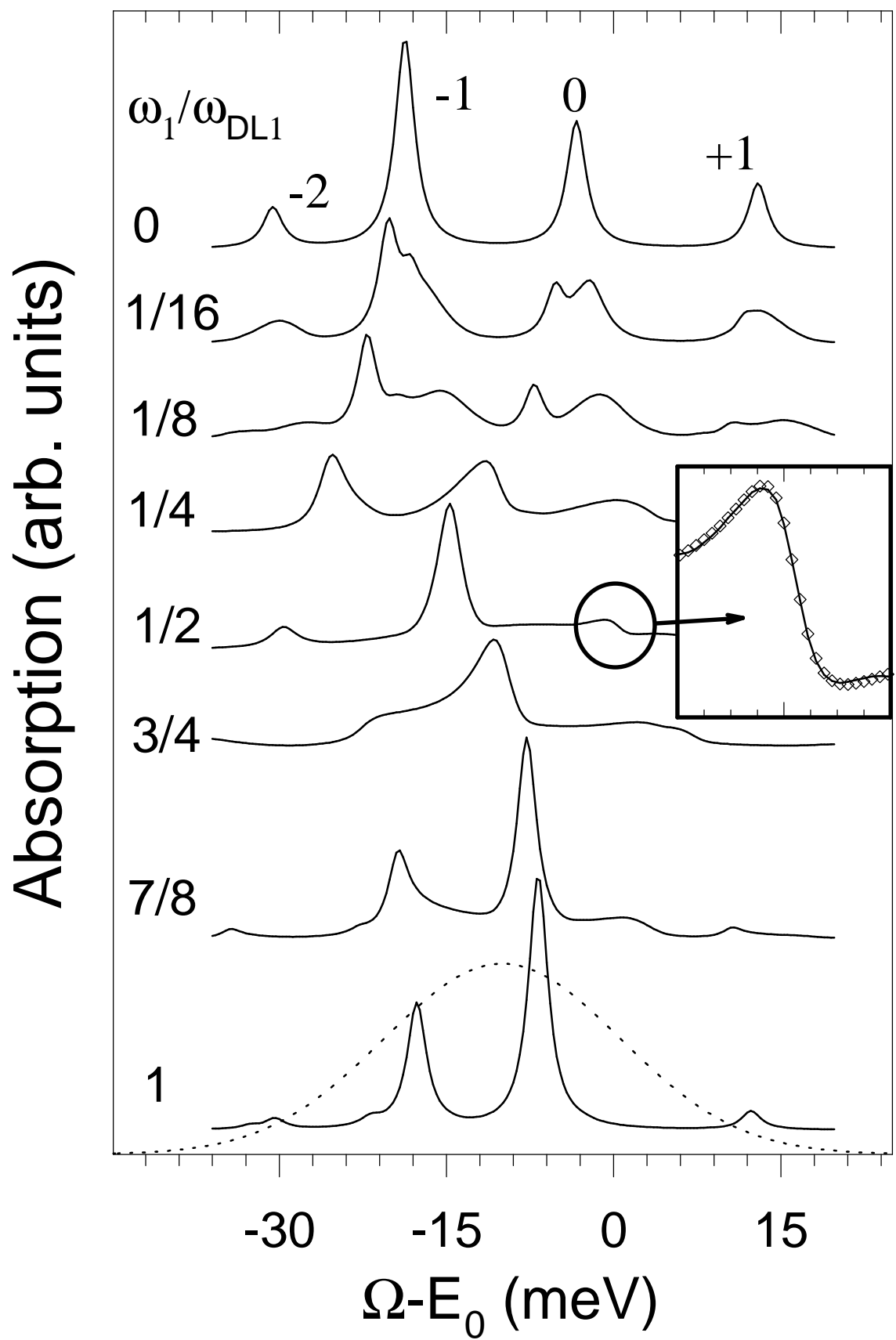
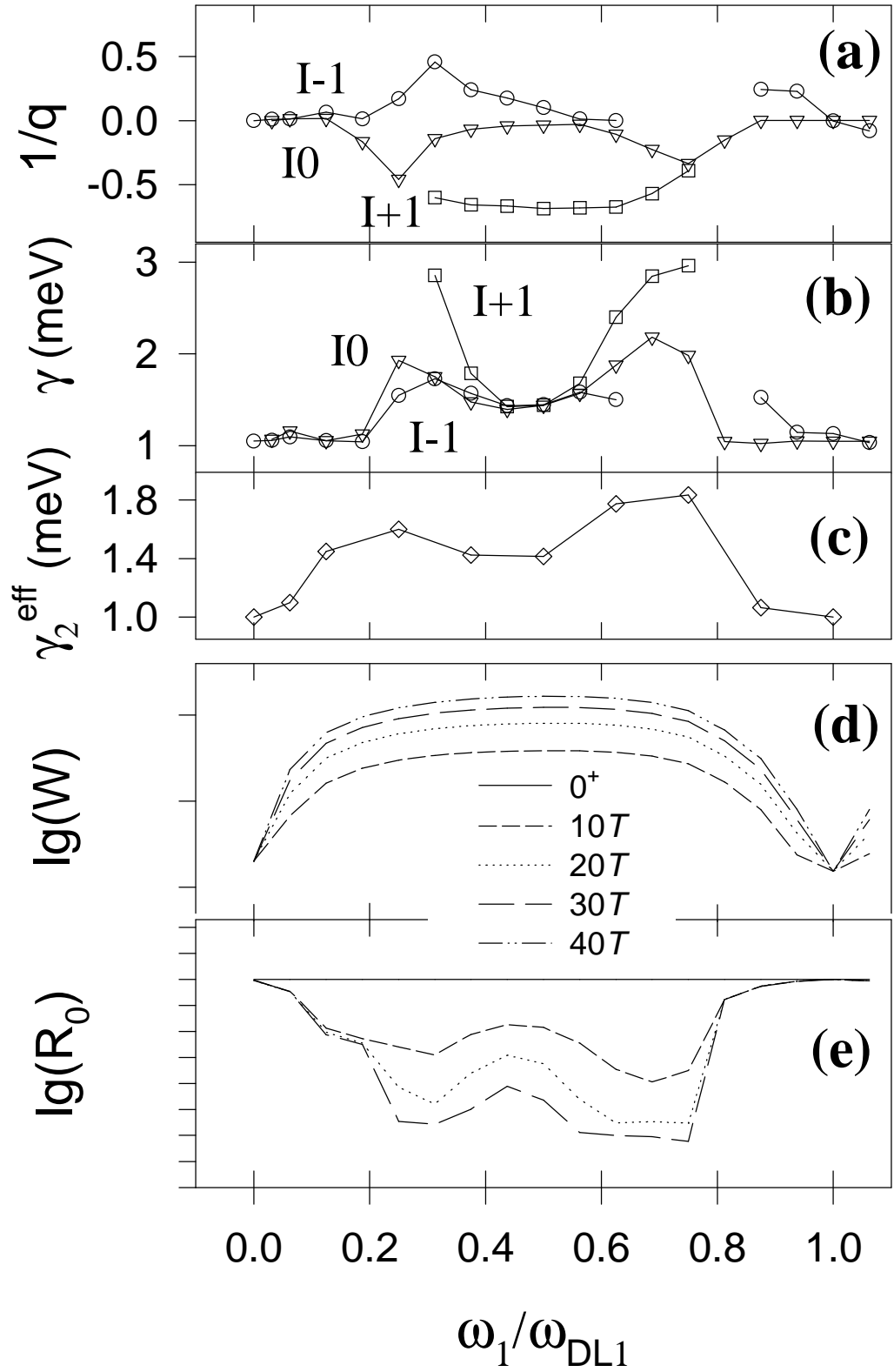
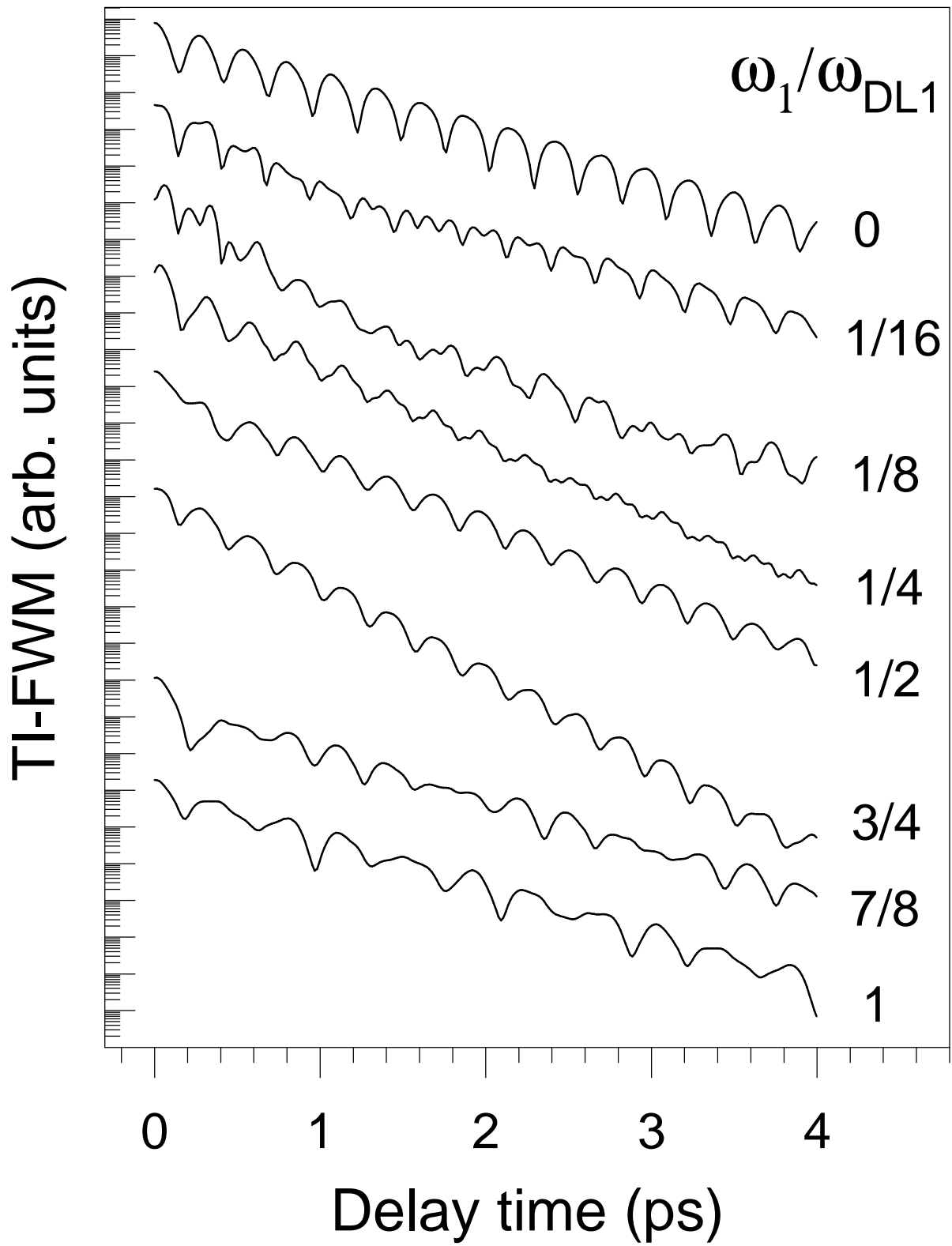
\end{document}